\documentstyle[12pt]{article}
\newlength{\pubnumber} 

\catcode`\@=11
\@addtoreset{equation}{section}

\def\section{\@startsection{section}{1}{\z@}{3.5ex plus 1ex minus .2ex}
 {2.3ex plus .2ex}{\large\bf}}
\def\subsection{\@startsection{subsection}{2}{\z@}{2.3ex plus .2ex}
 {2.3ex plus .2ex}{\bf}}

\def\LG{Landau--Ginzburg}
\begin{document}

\begin{titlepage}
\samepage{
\setcounter{page}{0}
\rightline{ACT--12/98}
\rightline{CERN--TH--98/395}
\rightline{CPT--TAMU--49/98}
\rightline{NSF--ITP--98--128}
\rightline{TPI--MINN--98/24}
\rightline{UFIFT-HEP-98--39}
\rightline{UMN--TH--1733--98}
\rightline{\tt hep-th/9812141}
\rightline{December 1998}
\vfill
\begin{center}
 {\Large \bf On Elevating Free-Fermion $Z_2\times Z_2$ Orbifolds  \\}
\vspace{.10in}
{\Large \bf Models to Compactifications of $F$ Theory}
\vfill
\vspace{.25in}
 {\large P. Berglund$^{1}$, J. Ellis$^{2}$,
   A.E. Faraggi$^{3}$, D.V. Nanopoulos$^{4}$,
   $\,$and$\,$ Z. Qiu$^5$\\}
\vspace{.25in}
{\it $^{1}$ Inst. for Theoretical Physics, University of California,
     Santa Barbara, CA~93106\\}
\vspace{.05in}
 {\it  $^{2}$ Theory Division, CERN, CH-1211 Geneva, Switzerland \\}
\vspace{.05in}
 {\it  $^{3}$ Department of Physics,
              University of Minnesota, Minneapolis, MN  55455, USA\\}
\vspace{.05in}
 {\it  $^{4}$ Dept. of Physics,
Texas A \& M University, College Station, TX~77843-4242, USA,  \\
HARC, The Mitchell Campus, Woodlands, TX~77381, USA, and \\
Academy of Athens, 28~Panepistimiou Avenue,
Athens 10679, Greece.\\}
\vspace{.05in}
 {\it  $^{5}$ Department of Physics,
              University of Florida, Gainesville, FL, 32611, USA\\}
\end{center}
\vfill
\begin{abstract}
  {\rm
We study the elliptic fibrations of some Calabi-Yau three-folds,
including the $Z_2\times Z_2$
orbifold with $(h_{1,1},h_{2,1})=(27,3)$, which is equivalent to
the common framework of realistic free-fermion models,
as well as related orbifold models with
$(h_{1,1},h_{2,1})=(51,3)$ and $(31,7)$. 
However,
two related puzzles arise when one considers  the 
$(h_{1,1},h_{2,1})=(27,3)$ model as
an $F$-theory compactification to six dimensions.
The condition for the vanishing of the gravitational
anomaly is not satisfied, suggesting that 
the $F$-theory compactification does not make sense, and
the elliptic fibration is well defined everywhere except at
four singular points in the base.
We speculate on the possible existence of $N=1$ tensor and
hypermultiplets at these points which would cancel the
gravitational anomaly in this case. }
\end{abstract}
\vfill
\smallskip}
\end{titlepage}

\setcounter{footnote}{0}

\def\beq{\begin{equation}}
\def\eeq{\end{equation}}
\def\beqn{\begin{eqnarray}}
\def\la{\label}
\def\eeqn{\end{eqnarray}}
\def\Tr{{\rm Tr}\,}
\def\KM{{Ka\v{c}-Moody}}

\def\ie{{\it i.e.}}
\def\etc{{\it etc}}
\def\eg{{\it e.g.}}
\def\half{{\textstyle{1\over 2}}}
\def\third{{\textstyle {1\over3}}}
\def\quarter{{\textstyle {1\over4}}}
\def\m{{\tt -}}
\def\p{{\tt +}}

\def\rep#1{{\bf {#1}}}
\def\slash#1{#1\hskip-6pt/\hskip6pt}
\def\slk{\slash{k}}
\def\GeV{\,{\rm GeV}}
\def\TeV{\,{\rm TeV}}
\def\y{\,{\rm y}}
\def\SM{Standard-Model }
\def\SUSY{supersymmetry }
\def\SSM{supersymmetric standard model}
\def\vev#1{\left\langle #1\right\rangle}
\def\l{\langle}
\def\r{\rangle}

\def\Htw{{\tilde H}}
\def\chibar{{\overline{\chi}}}
\def\qbar{{\overline{q}}}
\def\ibar{{\overline{\imath}}}
\def\jbar{{\overline{\jmath}}}
\def\Hbar{{\overline{H}}}
\def\Qbar{{\overline{Q}}}
\def\abar{{\overline{a}}}
\def\alphabar{{\overline{\alpha}}}
\def\betabar{{\overline{\beta}}}
\def\tautwo{{ \tau_2 }}
\def\calF{{\cal F}}
\def\calP{{\cal P}}
\def\calN{{\cal N}}
\def\smallmatrix#1#2#3#4{{ {{#1}~{#2}\choose{#3}~{#4}} }}
\def\bone{{\bf 1}}
\def\V{{\bf V}}
\def\N{{\bf N}}
\def\bQ{{\bf Q}}
\def\t#1#2{{ \Theta\left\lbrack \matrix{ {#1}\cr {#2}\cr }\right\rbrack }}
\def\C#1#2{{ C\left\lbrack \matrix{ {#1}\cr {#2}\cr }\right\rbrack }}
\def\tp#1#2{{ \Theta'\left\lbrack \matrix{ {#1}\cr {#2}\cr }\right\rbrack }}
\def\tpp#1#2{{ \Theta''\left\lbrack \matrix{ {#1}\cr {#2}\cr }\right\rbrack }}
\def\ul#1{$\underline{#1}$}
\def\bE#1{{E^{(#1)}}}
\def\IZ{\relax{\bf Z}}\def\IC{\relax{\bf C}}
\def\IR{\relax{\rm I\kern-.18em R}}
\def\lamb{\lambda}
\def\fc#1#2{{#1\over#2}}
\def\hx#1{{\hat{#1}}}
\def\Gh{\hat{\Gamma}}
\def\subsubsec#1{\noindent {\it #1} \br}
\def\WP{{\bf WP}}
\def\gn{\Gamma_0}
\def\bgn{{\bar \Gamma}_0}
\def\Ds{\Delta^\star}
\def\abstract#1{
\vskip .5in\vfil\centerline
{\bf Abstract}\penalty1000
{{\smallskip\ifx\answ\bigans\leftskip 2pc \rightskip 2pc
\else\leftskip 5pc \rightskip 5pc\fi
\noindent\abstractfont \baselineskip=12pt
{#1} \smallskip}}
\penalty-1000}
\def\us#1{\underline{#1}}
\def\hth/#1#2#3#4#5#6#7{{\tt hep-th/#1#2#3#4#5#6#7}}
\def\nup#1({Nucl.\ Phys.\ $\us {B#1}$\ (}
\def\plt#1({Phys.\ Lett.\ $\us  {B#1}$\ (}
\def\cmp#1({Commun.\ Math.\ Phys.\ $\us  {#1}$\ (}
\def\prp#1({Phys.\ Rep.\ $\us  {#1}$\ (}
\def\prl#1({Phys.\ Rev.\ Lett.\ $\us  {#1}$\ (}
\def\prv#1({Phys.\ Rev.\ $\us  {#1}$\ (}
\def\mpl#1({Mod.\ Phys.\ Let.\ $\us  {A#1}$\ (}
\def\ijmp#1({Int.\ J.\ Mod.\ Phys.\ $\us{A#1}$\ (}
\def\br{\hfill\break}\def\ni{\noindent}
\def\mbr{\hfill\break\vskip 0.2cm}
\def\cx#1{{\cal #1}}\def\al{\alpha}\def\IP{{\bf P}}
\def\ov#1#2{{#1 \over #2}}
\def\be{\beta}\def\al{\alpha}
\def\b{{\bf b}}
\def\S{{\bf S}}
\def\X{{\bf X}}
\def\I{{\bf I}}
\def\mb{{\mathbf b}}
\def\mS{{\mathbf S}}
\def\mX{{\mathbf X}}
\def\mI{{\mathbf I}}
\def\balpha{{\mathbf \alpha}}
\def\bbeta{{\mathbf \beta}}
\def\bgamma{{\mathbf \gamma}}
\def\bxi{{\mathbf \xi}}
 
\def\ul#1{$\underline{#1}$}
\def\bE#1{{E^{(#1)}}}
\def\IZ{\relax{\bf Z}}\def\IC{\relax{\bf C}}
\def\IR{\relax{\rm I\kern-.18em R}}
\def\lam{\lambda}
\def\fc#1#2{{#1\over#2}}
\def\hx#1{{\hat{#1}}}
\def\Gh{\hat{\Gamma}}
\def\subsubsec#1{\noindent {\it #1} \br}
\def\WP{{\bf WP}}
\def\gn{\Gamma_0}
\def\bgn{{\bar \Gamma}_0}
\def\Ds{\Delta^\star}
\def\abstract#1{
\vskip .5in\vfil\centerline
{\bf Abstract}\penalty1000
{{\smallskip\ifx\answ\bigans\leftskip 2pc \rightskip 2pc
\else\leftskip 5pc \rightskip 5pc\fi
\noindent\abstractfont \baselineskip=12pt
{#1} \smallskip}}
\penalty-1000}
\def\us#1{\underline{#1}}
\def\hth/#1#2#3#4#5#6#7{{\tt hep-th/#1#2#3#4#5#6#7}}
\def\nup#1({Nucl.\ Phys.\ $\us {B#1}$\ (}
\def\plt#1({Phys.\ Lett.\ $\us  {B#1}$\ (}
\def\cmp#1({Commun.\ Math.\ Phys.\ $\us  {#1}$\ (}
\def\prp#1({Phys.\ Rep.\ $\us  {#1}$\ (}
\def\prl#1({Phys.\ Rev.\ Lett.\ $\us  {#1}$\ (}
\def\prv#1({Phys.\ Rev.\ $\us  {#1}$\ (}
\def\mpl#1({Mod.\ Phys.\ Let.\ $\us  {A#1}$\ (}
\def\ijmp#1({Int.\ J.\ Mod.\ Phys.\ $\us{A#1}$\ (}
\def\br{\hfill\break}\def\ni{\noindent}
\def\mbr{\hfill\break\vskip 0.2cm}
\def\cx#1{{\cal #1}}\def\al{\alpha}\def\IP{{\bf P}}
\def\ov#1#2{{#1 \over #2}}
\def\be{\beta}\def\al{\alpha}


\def\inbar{\,\vrule height1.5ex width.4pt depth0pt}

\def\IC{\relax\hbox{$\inbar\kern-.3em{\rm C}$}}
\def\IQ{\relax\hbox{$\inbar\kern-.3em{\rm Q}$}}
\def\IR{\relax{\rm I\kern-.18em R}}
 \font\cmss=cmss10 \font\cmsss=cmss10 at 7pt
\def\IZ{\relax\ifmmode\mathchoice
 {\hbox{\cmss Z\kern-.4em Z}}{\hbox{\cmss Z\kern-.4em Z}}
 {\lower.9pt\hbox{\cmsss Z\kern-.4em Z}}
 {\lower1.2pt\hbox{\cmsss Z\kern-.4em Z}}\else{\cmss Z\kern-.4em Z}\fi}

\def\AEF{A.E. Faraggi}
\def\KRD{K.R. Dienes}
\def\JMR{J. March-Russell}
\def\NPB#1#2#3{{\it Nucl.\ Phys.}\/ {\bf B#1} (19#2) #3}
\def\PLB#1#2#3{{\it Phys.\ Lett.}\/ {\bf B#1} (19#2) #3}
\def\PRD#1#2#3{{\it Phys.\ Rev.}\/ {\bf D#1} (19#2) #3}
\def\PRL#1#2#3{{\it Phys.\ Rev.\ Lett.}\/ {\bf #1} (19#2) #3}
\def\PRT#1#2#3{{\it Phys.\ Rep.}\/ {\bf#1} (19#2) #3}
\def\MODA#1#2#3{{\it Mod.\ Phys.\ Lett.}\/ {\bf A#1} (19#2) #3}
\def\IJMP#1#2#3{{\it Int.\ J.\ Mod.\ Phys.}\/ {\bf A#1} (19#2) #3}
\def\nuvc#1#2#3{{\it Nuovo Cimento}\/ {\bf #1A} (#2) #3}
\def\etal{{\it et al,\/}\ }

\hyphenation{su-per-sym-met-ric non-su-per-sym-met-ric}
\hyphenation{space-time-super-sym-met-ric}
\hyphenation{mod-u-lar mod-u-lar--in-var-i-ant}


\setcounter{footnote}{0}
\section{Introduction}

Important progress has been achieved during the past few years in
understanding
non--perturbative aspects of superstring theories~\cite{reviewnps}.
However, the ultimate goal of understanding how string theory is relevant
to physics in the real world remains elusive. Encouraged by the hope
that string theory provides a framework for consistently
unifying all of the observed elementary
matter particles and interactions~\cite{initial}, many phenomenological
string models have been developed~\cite{reviewsp}. 
Among the most advanced
models are those constructed in the
free-fermion formulation~\cite{fsu5,fny,alr,eu,lykken}. These models
have been the subject of detailed studies showing that they
can, at least in
principle, account for desirable physical features 
including the observed fermion
mass spectrum, the longevity of the proton,
small neutrino masses,
the consistency of gauge-coupling unification
with the experimental data from LEP and elsewhere, and the 
universality of the soft supersymmetry-breaking
parameters~\cite{reviewsp}.
It is plausible that some improved understanding
how recent advances in non-perturbative
aspects of string theory are relevant in the real world
may  be gained by studying their application to these
realistic free-fermion models~\footnote{Other approaches are
also possible: see for example the $M$-theory three-generation
models proposed in~\cite{donagi}. It will be interesting
to see whether such models share the attractive features of
the perturbative three-generation models.}.

An important feature of the realistic free-fermion
models is their underlying $Z_2\times Z_2$
orbifold structure. Many of the encouraging phenomenological
characteristics of the realistic free-fermion
models are rooted in this
structure, including the three generations arising from the 
three twisted sectors, and the canonical $SO(10)$ embedding for
the weak hypercharge. To see more precisely this
orbifold correspondence, recall that 
the free-fermion models are generated by a set of basis vectors
which define the transformation properties of the world--sheet
fermions as they are transported around loops on the string world sheet.
A large set of realistic free-fermion models contains
a subset of boundary conditions which
can be seen
to correspond to $Z_2\times Z_2$ orbifold compactification
with the standard embedding of the gauge connection~\cite{foc}.
This underlying free-fermion model contains 24 generations from the
twisted sectors, as well as three additional 
generation/anti--generation pairs from the untwisted sector.
At the free-fermion point in the Narain moduli space~\cite{Narain},
both the metric and the antisymmetric background fields
are non-trivial, leading to an $SO(12)$ enhanced symmetry group.
The action of the $Z_2\times Z_2$ twisting on the
$SO(12)$ Narain lattice then gives rise to a model
with $(h_{11},h_{21})=(27,3)$, matching the data of
the free-fermion model~\footnote{We
emphasize that the data of this 
model differs from the $Z_2\times Z_2$ orbifold on a $SO(4)^3$
lattice with $(h_{11},h_{21})=(51,3)$, which has been
more extensively discussed in the literature \cite{z2mf}.}.

Recently, we have shown how to construct this
$Z_2\times Z_2$ orbifold model 
in the Landau--Ginzburg formalism~\cite{befnq}. 
This was done using a freely-acting $Z_2$ twist on a 
$Z_2\times Z_2$ Landau-Ginzburg orbifold with
$(h_{11},h_{21})=(51,3)$. 
In this paper, we extend this
analysis to include a formulation of this 
and related $(51,3)$ and $(31,7)$ models in terms of
elliptically-fibered Calabi-Yau manifolds,
opening the way to non-trivial $F$- or $M$-theory
compactifications. This geometrical approach should allow us to
study properties of the moduli space away from the special
free-fermion/orbifold point. However, in this paper
we are more concerned with the consistency
of these manifolds as valid $F$-theory
compactifications, and thus as six-dimensional vacua, rather
than exploring them from the traditional heterotic four-dimensional point
of view.
 
We recall that
$F$ theory is a way of compactifying type-IIB string theory
which allows the string coupling to vary over the compact
manifold. 
The key point in compactifications of $F$ theory to six
dimensions is that
the models should admit an elliptic fibration and (at least) a global
section,  in which a Calabi--Yau
three-fold is identified as a two complex--dimensional base
manifold $B$ with an elliptic fiber. Among these
models are some which have an orbifold interpretation, such as
the above-mentioned $Z_2\times Z_2$ orbifold
with $(h_{11},h_{21})=(51,3)$~\cite{z2mf}, denoted by $X_1$,
as we demonstrate with a standard Weierstrass representation. 
Using the Landau-Ginzburg analysis and {\it quartic} 
Weierstrass representations, we construct related
freely-acting $Z_2$ orbifolds with $(h_{11},h_{21})=(27,3)$
and $(h_{11},h_{21})=(31,7)$, 
denoted by $X_2$ and $X_3$. The former admits an
elliptic fibration, apart from four singular
points in the base
$B$. However, these points prevent us from having a global
section and so the six-dimensional theory does not exist.
Another sign of this is that
the gravitational anomaly equation in six dimensions is not
satisfied. This implies that the formulation of this $F$-theory 
compactification is not well defined. This is consistent with
the absence of a globally-defined section, but it is possible that 
there may be a non-trivial contribution of $N=1$
tensor and hypermultiplets associated with the singular points in the base 
$B$ of $X_2$,
due to the $Z_2$ quotient, which cancels the gravitational anomaly.
On the other hand, we are able to show that the $(31,7)$ model $X_3$ does 
admit a consistent elliptic fibration.

This paper is organized as follows. In section 2, we give a brief review
of the free-fermion orbifold and Landau-Ginzburg constructions of the
$(h_{11},h_{21})=\{(51,3),(27,3)\}$ $Z_2\times Z_2$ orbifolds. In section
3, we first review the elliptic fibration of the former orbifold
using a standard Weierstrass representation, and then
generalize it to the closely related $(27,3)$ and $(31,7)$ models
using a quartic Weierstrass representation. In particular, we
focus on the question of the existence of the
$(27,3)$ model in six dimensions and
the appearance of singular points in the base space, and suggest how the
issue of the gravitational anomaly may be resolved. We study in section 4
type-IIB orientifold constructions relevant for the discussion of the
$F$-theory compactification of $X_2$. Finally, we end in section 5 with
some concluding remarks. 

\setcounter{footnote}{0}
\section{The $Z_2\times Z_2$ Orbifold Equivalent of Realistic
Free-Fermion Models}

The purpose of this section is to motivate the choice of the relevant
Calabi-Yau
three-folds because of their relation to certain free-fermionic models. We
first review the construction of
the $Z_2\times Z_2$ fermionic orbifold of interest followed by their
realization as Landau-Ginzburg and toroidal orbifolds.

Let us recall that,
in the free-fermion formulation~\cite{fff}, a model
is defined by a set of boundary-condition basis
vectors, together with the related one--loop
GSO-projection coefficients, that are constrained
by the string consistency constraints.
These boundary-condition basis vectors encode the phases
of all the world--sheet fermions, when transported
around one of the non-contractible loops of the
string world sheet. In the case of the heterotic string
in the light--cone gauge, there are 20 left--moving and
44 right--moving real Majorana--Weyl world--sheet fermions,
whereas for the type-IIA and type-IIB strings there would be
20 left-- and 20 right--moving world--sheet fermions.
Given the set of boundary-condition basis vectors and the one--loop
GSO-projection coefficients, one can then construct the one--loop
partition function and extract the physical spectrum.

The $Z_2\times Z_2$ fermionic orbifold model 
of interest is generated
by the following set of boundary-condition basis vectors, the 
so-called NAHE set~\cite{fsu5,nahe}:
$
\{{\bf 1},\S,\bxi={\bf 1}+\b_1+\b_2+\b_3,X,\b_1,\b_2\},
$
The first four vectors
in the basis $\{{\bf 1},\S,\bxi,\X\}$ generate a model with $N=4$
space--time supersymmetry, and an $E_8\times SO(12)\times E_8$ gauge
group. In this construction,
the sector {\bf \S} generates  $N=4$ space--time supersymmetry.
The $SO(12)$ factor is obtained
from $\{{\bar y},{\bar\omega}\}^{1,\cdots,6}$.
The first and second $E_8$ factors are obtained from the world--sheet
fermionic states $\{{\bar\psi}^{1,\cdots,5},{\bar\eta}^{1,2,3}\}$ and
$\{{\bar\phi}^{1,\cdots,8}\}$, respectively.
The sectors $\X$ and
$\bxi$ produce the spinorial representations of $SO(16)$ in the
observable and hidden sectors, respectively,
and complete the observable and hidden gauge groups to $E_8\times E_8$.
The Neveu--Schwarz sector
produces the adjoint representations of $SO(16)\times SO(12)\times SO(16)$.
The vectors $\b_1$ and $\b_2$ break the gauge symmetry to
$E_6\times U(1)^2\times SO(4)^3\times E_8$ and 
the $N=4$ space--time supersymmetry to $N=1$.
The sectors $(\b_1;\b_1+\X)$, $(\b_2;\b_2+\X)$ and $(\b_3;\b_3+\X)$
each give eight ${\bf 27}$'s of $E_6$. The $(NS;NS+\X)$ sector gives, in
addition to the vector bosons and spin-two states, three copies of
scalar representations in ${\bf 27}+{\overline{\bf 27}}$ of $E_6$.
The net number of generations 
in the {\bf 27} representation of $E_6$
is therefore 24.

In the toroidal orbifold construction, the same model is obtained
by first specifying the background fields, which produce
the $SO(12)$ lattice~\cite{foc}, and then
applying the appropriate $Z_2\times Z_2$ identifications.
One takes the metric
on the six-dimensional compactified manifold
to be the Cartan matrix of $SO(12)$,
and the antisymmetric tensor to be $b_{ij}=g_{ij}$ for $i>j$ \cite{foc}.
When all the radii of the six-dimensional compactified
manifold are fixed at $R_I=\sqrt2$, it is easily seen that the
left-- and right--moving momenta
$P^I_{R,L}=[m_i-{1\over2}(B_{ij}{\pm}G_{ij})n_j]{e_i^I}^*$
reproduce all the massless root vectors in the lattice of
$SO(12)$,
where the $e^i=\{e_i^I\}$ are six linearly-independent
vectors normalized: $(e_i)^2=2$.
The ${e_i^I}^*$ are dual to the $e_i$, and
$e_i^*\cdot e_j=\delta_{ij}$. The momenta $P^I$ of the compactified
scalars in the bosonic formulation can be seen to coincide
with the $U(1)$ charges of the unbroken Cartan generators
of the four-dimensional gauge group.

The incorporation in the free-fermion model of
the two basis vectors $\b_1$ and $\b_2$ as well as
$\{{\bf 1},\S,\bxi,\X\}$
corresponds to the $Z_2\times Z_2$
orbifold model with standard embedding.
The fermionic boundary conditions are translated,
in the bosonic language, into twists on the internal dimensions
and shifts in the gauge degrees of freedom.
Starting from the model with $SO(12)\times E_8\times E_8$
symmetry, and applying the $Z_2\times Z_2$ twisting on the
internal
coordinates, we then obtain an orbifold model with $SO(4)^3\times
E_6\times U(1)^2\times E_8$ gauge symmetry. There are sixteen fixed
points in each twisted sector, yielding the 24 generations from the
three twisted sectors mentioned above. The three additional pairs of
${\bf 27}$
and ${\bf \overline{27}}$
are obtained from the untwisted sector. This
orbifold model, which we call $X_2$, therefore has the same
topological data as the free-fermion model
with the six-dimensional basis set
$\{{\bf1},\S,\X,\I={\bf1}+\b_1+\b_2+\b_3,\b_1,\b_2\}$, since
the Euler characteristic of this model is 48, with $h_{11}=27$ and
$h_{21}=3$.

This $Z_2\times Z_2$ orbifold, corresponding
to the extended NAHE set at the core of the realistic
free--fermion models,
differs from the one which has usually been
examined in the literature~\cite{z2mf}.
In that orbifold model, the Narain
lattice is $SO(4)^3$, yielding a $Z_2\times Z_2$ orbifold model, 
which we call $X_1$.
It has Euler characteristic equal to 96, corresponding to 48 generations,
and $h_{11}=51$, $h_{21}=3$. This $Z_2\times Z_2$ orbifold
can be constructed in a similar manner to the model $X_2$ above. 
First the Narain $SO(4)^3$
lattice is produced via the diagonal metric $g_{ij}=2\delta_{ij}$
and the trivial anti-symmetric tensor field $b_{ij}=0$.
For $R_{I}=\sqrt{2}$, all the roots in the root
lattice of $SO(4)^3$ are again massless. Then, applying
the $Z_2\times Z_2$ twisting reduces the $N=4$ supersymmetry
to $N=1$. Each twisted sector now produces
16 generations, yielding a total of 48, and three
additional generation and anti-generation pairs
are obtained from the untwisted sector.

Before proceeding, we note that, at the level of the toroidal
compactification, the $SO(4)^3$ and $SO(12)$ lattices
are continuously connected by varying the parameters of the
background metric and antisymmetric tensor.
However, this cannot be done while preserving the
$Z_2\times Z_2$ invariance, because 
the continuous interpolation cannot change the Euler characteristic.
Therefore, the two toroidal models are in the same
moduli space, but not the two orbifold models $X_1$ and $X_2$.

Let us now show how the two $Z_2\times Z_2$
orbifold models, constructed above using toroidal
compactification, may be
represented  in the Landau-Ginzburg orbifold 
construction~\cite{befnq}. We start from  a non-degenerate
quasi--homogeneous superpotential $W$ of degree $d$,
$W(\lambda^{n_i}X_i)=\lambda^d W(X_i)$,
where the $X_i$ are chiral superfields and the $q_i=n_i/d$ are their
left and right charges under the $U(1)_{J_0}$ current
of the $N=2$ algebra. In the \LG~ construction
one twists by some symmetry group $G$
of the original superpotential \cite{lg}. The \LG~ potential
that mimics the $T_2^3$ torus, corresponding to the $SO(4)^3$ lattice,
 is given by
\beq
W=X_1^4+X_2^4+X_3^2+X_4^4+X_5^4+X_6^2+X_7^4+X_8^4+X_9^2
\label{quarticpotential}
\eeq
where $X_{3,6,9}$ are trivial superfields, and  
the superpotential (\ref{quarticpotential})
corresponds to a superconformal field theory with
$c=9$. The mirror of the $X_1$ model is obtained
by taking the orbifold
${\cal M}/(Z_2^A\times Z_2^B)$ where
\beqn
&& Z_2^A~:~(X_1,\cdots,X_9)~\rightarrow~ 
   (X_1, X_2,X_3, -X_4, -X_5,X_6,-X_7,-X_8,X_9)~;~~~~~~~~~~\nonumber\\
&& Z_2^B~:~(X_1,\cdots,X_9)~\rightarrow~ 
   (-X_1, -X_2, X_3, -X_4, -X_5,X_6,X_7,X_8,X_9)~.~~~~~~~~~~\label{z2z2513}
\eeqn
and ${\cal M}=W/j$, where $j$ is the $Z_4$ scaling symmetry
of~(\ref{quarticpotential}). 
It is easy to show that there are 51 $(1,1)$ and 3 $(-1,1)$
states. Using the convention that the deformations of $W$, which give
part of the spectrum of $(1,1)$ states, correspond to the
$(2,1)$ forms of the orbifold, and the fact that the $(1,1)$ forms are in
one-to-one correspondence
with the $(-1,1)$ states, we reproduce the data of the $Z_2\times Z_2$
orbifold on the $SO(4)^3$ lattice. 
We showed in~\cite{befnq} that the mirror of the $X_2$ model is
obtained from the mirror of the $X_1$ model
by applying the twist
\beqn
&& Z_2^{w}~:~(X_1,\ldots, X_9)~\rightarrow~ 
   (-X_1, X_2, -X_3, -X_4, X_5, -X_6, -X_7, X_8, -X_9)~,~~~~~~~
                                        \label{z2z2273}
\eeqn
i.e., we have the \LG\ orbifold ${\cal M}/(Z_2^A\times Z_2^B\times
Z_2^\omega)$.
Note that the three trivial superfields are twisted by the
$Z_2^w$ twist. This is to ensure that $Z_2^w$ acts freely on each of
the $T^2$ factors in~(\ref{quarticpotential}),
thus reproducing the data of the $Z_2\times Z_2$ orbifold
on the $SO(12)$ lattice.

Another way to realize the connection between the $X_1$ and $X_2$ 
models is by using a freely-acting shift, rather than a freely-acting
twist~\footnote{This second way will be instrumental later in trying
to find a
type-IIB orientifold
description of the six-dimensional vacuum corresponding to the $F$-theory
compactification on $X_2$.}.
For this purpose, let us first start with the compactified
$T^2_1\times T^2_2\times T^2_3$ torus parameterized by
three complex coordinates $z_1$, $z_2$ and $z_3$,
with the identification
\beq
z_i=z_i + 1~~~~~~~~~~;~~~~~~~~~~z_i=z_i+\tau_i
\label{t2cube}
\eeq
where $\tau$ is the complex parameter of each
$T^2$ torus. 
We consider $Z_2$ twists and possible shifts of order
two:
\beq
z_i~\rightarrow~(-1)^{\epsilon_i}z_i+1/2\delta_i
\la{z2twistanddance}
\eeq
subject to the condition that $\Pi_i(-1)^{\epsilon_i}=1$.
This condition insures that the holomorphic three--form
$\omega=dz_1\wedge z_2\wedge z_3$ is invariant under the $Z_2$ twist.
Under the identification $z_i\rightarrow-z_i$, a single torus 
has four fixed points at
\beq
z_i=\{0,1/2,\tau/2,(1+\tau)/2\}.
\la{fixedtau}
\eeq
The first model that we consider is produced
by the two $Z_2$ twists 
\beqn
&& \alpha:(z_1,z_2,z_3)\rightarrow(-z_1,-z_2,~~z_3)\cr
&&  \beta:(z_1,z_2,z_3)\rightarrow(~~z_1,-z_2,-z_3)
\la{alphabeta}
\eeqn
There are three twisted sectors in this model, $\alpha$,
$\beta$ and $\alpha\beta=\alpha\cdot\beta$, each producing
16 fixed tori, for a total of 48. 

To facilitate the
discussion of the subsequent examples, we briefly describe
the calculation of the cohomology for this orbifold:
a more complete discussion can be found in~\cite{msanddt}.
Consider first the untwisted sector. The Hodge
diamond for a single untwisted torus is given by
\beq
\left(\matrix{1 &1\cr
              1 &1\cr
             } \right)
\la{hodge1t}
\eeq
which displays the dimensions of the $H^{p,q}(T_i)$,
with $H^{0,0}$, $H^{0,1}$, $H^{1,0}$ and $H^{1,1}$
being generated by the differential forms $1$, $d{\bar z}_i$,
$dz_i$ and $dz_i\wedge d{\bar z}i$, respectively. Under the $Z_2$
transformation $z~\rightarrow~-z$, $H^{0,0}$ and $H^{1,1}$ 
are invariant, whereas $H^{1,0}$ and $H^{0,1}$ change sign.

The untwisted sector of the manifold produced
by the product of the three tori $T_1\times T_2\times T_3$
is then given by the product of differential forms
which are invariant under the $Z_2\times Z_2$ twists
$\alpha\times \beta$. The invariant terms
are summarized by the Hodge diamond
\beq
\left(\matrix{ 1&0&0&1\cr
               0&3&3&0\cr
               0&3&3&0\cr
               1&0&0&1\cr}\right)
\la{hodge3t}
\eeq
For example, $H^{1,1}$ is generated by $dz_i\wedge {\bar z}_i$ 
for $i=1,2,3$, and $H^{2,1}$ is produced by 
$dz_1\wedge z_2\wedge{\bar z}_3$, $dz_2\wedge z_3\wedge{\bar z}_1$,
$dz_3\wedge z_1\wedge{\bar z}_2$, etc..
We next turn to the twisted sectors, of which there are three,
produced by $\alpha$, $\beta$ and $\alpha\beta$, respectively.
In each sector, two of the $z_i$ are identified under $z_i\rightarrow-z_i$,
and one torus is left invariant. We need then consider
only one of the twisted sectors, say $\alpha$, and the others
will contribute similarly. The sector $\alpha$ has 16 fixed
points from the action of the twist on the first and second
tori. Since the action is trivial on the third torus, we get 16 fixed
tori. The cohomology
is given by sixteen copies of the cohomology of $T_3$,
where each $H^{p,q}$ of $T_3$ contributes $H^{p+1,q+1}$
to that of the orbifold theory~\cite{msanddt}.
The Hodge diamond from each twisted sector then 
has the form
\beq
\left(\matrix{ 0& 0& 0&0\cr
               0&16&16&0\cr
               0&16&16&0\cr
               0& 0& 0&0\cr}\right)
\la{hodgets}
\eeq
It remains to find the forms from the $\alpha$ twisted sector
which are invariant under the action of the $\beta$ twist. 
Since $z_3\rightarrow-z_3$ under $\beta$, it follows that 
$1$ and $dz_3\wedge d{\bar z}_3$ are invariant,
whereas $dz_3$ and $d{\bar z}_3$ are not.
Consequently, only the contributions of $H^{1,1}$
and $H^{2,2}$ in (\ref{hodgets}) are invariant 
under the $\beta$ twist. 
Therefore, we see that the invariant contribution
from each twisted sector is only along the diagonal of
(\ref{hodgets}), and that the total Hodge diamond of the
$Z_2\times Z_2$ orbifold is 
\beq
\left(\matrix{ 1& 0& 0&1\cr
               0&51& 3&0\cr
               0& 3&51&0\cr
               1& 0& 0&1\cr}\right)
\la{hodgez2z2}
\eeq

Next we consider the model generated by the $Z_2\times Z_2$
twists in (\ref{alphabeta}), with the additional shift
\beq
\gamma:(z_1,z_2,z_3)\rightarrow(z_1+{1\over2},z_2+{1\over2},z_3+{1\over2})
\la{gammashift}
\eeq
This model again has fixed tori from the three
twisted sectors $\alpha$, $\beta$ and $\alpha\beta$.
The product of the $\gamma$ shift in (\ref{gammashift})
with any of the twisted sectors does not produce any additional
fixed tori. Therefore, this shift acts freely.
Under the action of the $\gamma$ shift, half
the fixed tori from each twisted sector are paired.
Therefore, the action of this shift is to reduce
the total number of fixed tori from the twisted sectors
by a factor of $1/2$. Consequently, the Hodge diamond for this model is
\beq
\left(\matrix{ 1& 0& 0&1\cr
               0&27& 3&0\cr
               0& 3&27&0\cr
               1& 0& 0&1\cr}\right)
\la{hodgez2z2shift3}
\eeq
with $(h_{11},h_{21})=(27,3)$. This model therefore
reproduces the data of the $Z_2\times Z_2$ orbifold
at the free-fermion point in the Narain moduli space.
The action of the freely-acting shift (\ref{gammashift})
is seen to be identical to the action of the freely-acting twist
(\ref{z2z2273}) that connects the (51,3) and (27,3)
models in the Landau-Ginzburg representation.

Finally, let us consider the model generated by the
twists (\ref{alphabeta}) with the additional
shift given by
\beq
\gamma:(z_1,z_2,z_3)\rightarrow(z_1+{1\over2},z_2+{1\over2},z_3)
\la{gammashift2}
\eeq
This model, denoted by $X_3$, again has three twisted sectors $\alpha$,
$\beta$ and $\alpha\beta$. Under the action of the $\gamma$ shift,
half of the fixed tori from these twisted
sectors are identified. These twisted sectors
therefore contribute to the Hodge diamond 
as in the previous model. However, the $\gamma$ shift
in (\ref{gammashift2}) does not act freely,
as its combination with $\alpha$ produces additional 
fixed tori, since, under the action of the product
$\alpha\cdot\gamma$, we have
\beq
\alpha\gamma:(z_1,z_2,z_3)\rightarrow(-z_1+{1\over2},-z_2+{1\over2},z_3)
\la{alphagamma}
\eeq
This sector therefore has 16 additional fixed tori.
Repeating the analysis as in the previous cases,
we see that, under the identification imposed by the $\alpha$ and $\beta$ 
twists, the invariant states from this sector
give rise to four additional (1,1) forms and four additional
(2,1) forms. The Hodge diamond for this model
therefore has the form
\beq
\left(\matrix{ 1& 0& 0&1\cr
               0&31& 7&0\cr
               0& 7&31&0\cr
               1& 0& 0&1\cr}\right)
\la{hodgez2z2shift4}
\eeq
with $(h_{11},h_{21})=(31,7)$.

As we discuss in subsequent sections,
the various $Z_2\times Z_2$ orbifold models 
discussed in this section display
interesting features when one considers
the possibility of elliptic fibration in the context of $F$ theory.

\setcounter{footnote}{0}
\section{Elliptic Fibration and $F$ Theory}

Let us now study compactifications of $F$ theory
on the different $Z_2\times Z_2$ orbifold models
analyzed in the previous section, in particular on $X_2$.
Our strategy is to study first the Weierstrass representation
of the elliptic fibration of the $F$-theory compactification on
the $Z_2\times Z_2$ orbifold $X_1$,
and then implement the extra twist, so as to obtain the $X_2$ orbifold. 
Finally, we discuss how one may hope to resolve the puzzle of the
non-vanishing gravitational anomaly that we advertised earlier.

When compactifying $F$ theory on a Calabi-Yau threefold, it is essential
that it should admit an elliptic fibration, with
base $B$ and a toroidal global section. Elliptically-fibered manifolds
are conveniently parameterized by writing the equation
for the toroidal fiber in the standard Weierstrass form:
\beq
y^2=x^3+f x + g
\la{weierform}
\eeq
which expresses the torus as a double cover of the complex plane
with three finite branch points and one branch point at infinity.
The functions $f$ and $g$ are polynomials of degree 8 and 12, respectively,
in the base coordinates. Compactifying on a Calabi-Yau three-fold,
Morrison and Vafa~\cite{vafamori,vafamorii} have shown 
that, in terms of the $h_{1,1}(B)$ of the base manifold
and the $h_{1,1}(X)$ and $h_{2,1}(X)$ of the Calabi-Yau three-fold $X$,
the number of neutral hypermultiplets is given by
\beq
H^0=h_{2,1}(X)+1 ~~,
\la{nhyper}
\eeq
the number of tensor multiplets is given by
\beq
T=h_{1,1}(B)-1~~,
\la{tensorhype}
\eeq
and the rank of the vector multiplets is given by
\beq
r(V)=h_{1,1}(X)-h_{1,1}(B) -1.
\la{rankgg}
\eeq
Finally, cancelation of the gravitational anomaly
in $N=1$ supergravity in six dimensions requires
the following relation between the numbers of neutral hyper-,
vector, and tensor multiplets \cite{d6anomaly}:
\beq
H^0-V=273-29 T.
\la{ganomaly}
\eeq
Although we will mainly be using the 
Weierstrass parameterization above, this is not so
convenient for some of the 
models written as orbifolds of toroidal compactifications. 
In these cases, it is sometimes easier to work with the
corresponding orientifold model, where such a model 
has been identified. Specifically,
only some of the orbifolds
studied in the previous section
have been shown to admit an elliptic
fibration,
and have already been discussed
in the context of $F$-theory compactification
to six dimensions~\cite{vafamorii}. 
These models are the special classes
of Calabi--Yau three-folds that have been analyzed by
Voisin~\cite{voisin} and Borcea~\cite{borcea}.
They have been further classified by
Nikulin \cite{nikulin} in terms of three invariants $(r,a,\delta)$.
In the context of our discussion, we note that both the $X_1$
and $X_3$ models
are part of this classification, whereas $X_2$ is not. 

Returning to the Weierstrass representation~(\ref{weierform}),
we consider an
elliptically-fibered Calabi-Yau manifold $X$ with base $\IP^1\times
\IP^1$. The $Z_2\times Z_2$ orbifold $X_1$ can then be realized
as a singular limit of $X$.
We can represent $X$ in Weierstrass form, as a (singular)
elliptic fiber which depends on the (inhomogeneous) coordinates
$w,\tilde w$ of the respective $\IP^1$:
\beq
y^2 = x^3 + f_8 (w,\tilde w) x z^4 + g_{12}(w,\tilde w) z^6~.
\la{weier}
\eeq
Here $f_8$ and $g_{12}$ are of bidegree eight and twelve,
respectively, in $w,\tilde w$.
This model has $h_{1,1}=3$ and $h_{2,1}=243$, and Sen~\cite{sengp} has
shown that, considered as an $F$-theory vacuum in six dimensions, 
it is equivalent to the
Gimon-Polchinski type-IIB orientifold on $T^4/Z_2$~\cite{gp}.

The next step is to choose
a particular complex structure for $f_8$ and $g_{12}$. 
To do so, we let
\beq
f_8 = \eta - 3 h^2\,,\quad
g_{12}=h(\eta - 2 h^2)~.
\la{singular}
\eeq
which implies that the discriminant, $\Delta = 4f^3 + 27 g^2$, takes
the form
\beq
\Delta = \eta^2 (4\eta-9h^2)~.
\la{disc}
\eeq
We then further restrict $f_8$ and $g_{12}$ by setting
\beq
h = K \prod_{i,j=1}^4(w- w_i)(\tilde w - \tilde w_j)\,,\quad
\eta = C \prod_{i,j=1}^4(w- w_i)^2(\tilde w - \tilde w_j)^2~.
\la{etah}
\eeq
Thus, as we approach any of $w=w_i$ (or $\tilde w=\tilde w_j$) we have
a $D_4$ singular fiber. This follows from Kodaira's classification of
ADE singularities~\cite{vafamori,vafamorii,bv}, and has
\beq\la{dfour}
f_8\sim (w-w_i)^2\,,\quad g_{12}\sim (w-w_i)^3\,,\quad \Delta\sim
(w-w_i)^6~.
\eeq
Thus, we have an enhanced $SO(8)^8$ gauge symmetry, since 
$i,j=1,\ldots,4$.

These $D_4$ singularities intersect in 16 points, $(w_i,\tilde w_j),\,
i,j=1,\ldots 4$ in the base $\IP^1\times \IP^1$. Due to the severity
of the individual singularities, at each point of intersection 
 one has to resolve the
base~\cite{aspinwall}, as follows.
At each point of intersection, we
have the following singular behavior:
\beq\la{dfourinter}
f_8\sim (w-w_i)^2(\tilde w-\tilde w_j)^2\,,\quad g_{12}\sim
(w-w_i)^3(\tilde w-\tilde w_j)^3\,,\quad \Delta\sim
(w-w_i)^6(\tilde w-\tilde w_j)^6~.
\eeq
Thus the order of the singularity is $(4,6,12)$ respectively for
$f_8,g_{12},\Delta$. However, just resolving the singular fiber is not
enough. We also have to blow the base up once at each of $(w_i,\tilde
w_j),\, i,j=1,\ldots 4$. Thus, in addition to the
enhanced gauge symmetry, we also obtain 16 additional tensor 
multiplets~\cite{vafamori,vafamorii}.

The resulting Calabi-Yau manifold 
yields
an $F$-theory compactification on the elliptically-fibered
Calabi--Yau three-fold corresponding to the $Z_2\times Z_2$ (51,3)
orbifold model $X_1$. To see that it has
$h_{1,1}=51$ and $h_{2,1}=3$, we first note that 
blowing up the base gives 16 
$(1,1)$ forms, and recall that the $SO(8)^8$ group has rank 32, which
contributes
another 32 $(1,1)$ forms to $h_{1,1}$. In addition, out of the original
250 deformations encoded in $f_8,g_{12}$ ($9^2+13^2$), we are left with
$C, K$ and $w_i,\tilde w_i,i=1,\ldots,4$. This leaves us with three
independent deformations, once we have used the $SL(2,\IC)$
reparametrization
of each of the $\IP^1$, as well as an overall rescaling
of~(\ref{weier}). (In
this way, we can fix three of the $w_i,\tilde w_i$ as well as $K$.)
{}From~(\ref{nhyper}), (\ref{tensorhype}), and (\ref{rankgg})
we then find that this six-dimensional $F$-theory vacuum has
$V=224$, $T=17$ and $H^0=4$~\cite{vafamorii}, which is consistent with
the formula (\ref{ganomaly}) for the vanishing of the gravitational anomaly.

To seek the elliptic fibration of the (27,3) orbifold model,
we have to implement the final $Z_2$ which acts as a 
freely-acting twist in the Landau--Ginzburg representation of the model,
or as a freely-acting shift as in~(\ref{gammashift}).  
It is obvious that the model written in the
form~(\ref{weier}) is not the correct way of representing the covering space
of the final $Z_2$. Rather, we have to rewrite~(\ref{weier}) in terms of a
quartic polynomial~\footnote{This form has appeared before in various 
contexts in F-theory, see e.g. \cite{aspgross,ibanez,kmv,cpr,bps,bkmt}},
\beq\la{weierfour}
\hat y^2 = \hat x^4 + \hat x^2 \hat z_2 \hat f_4 + \hat x z^3 \hat g_6
+ \hat z^4 \hat h_8~.
\eeq
where $\hat f_4,\hat g_6,\hat h_8$ are of bidegree $4,6,8$
respectively in $w,\tilde w$. The relation between~(\ref{weier}) and
(\ref{weierfour}) is given in terms of 
\beq\la{rel}
\hat f_4 = -3 h\,,\quad \hat g_6 = 0\,,\quad \hat h_8 = -1/4 \eta~.
\eeq
Writing the fibered torus in the quartic form~(\ref{weierfour})
amounts to bringing the branch point at infinity to a finite
point. The reason for this rewriting of the fibered
torus is that in this representation some
symmetries of the torus become manifest, thus simplifying
the analysis.

We now note that~(\ref{weierfour}) enjoys a $Z_2$ symmetry: $(\hat y,\hat
x, \hat z)\to (-\hat y,-\hat x, \hat z)$~\cite{bps,bkmt}. 
We are, however, interested in an action which extends to the base as
well:
$(\hat y,\hat x, \hat z,w,\tilde w)\to (-\hat y,-\hat x, \hat
z,-w,-\tilde w)$. In order to carry out this identification, we need to
modify~(\ref{etah}) to
\beq\la{etahztwo}
h = K \prod_{i,j=1}^2(w^2- w_i^2)(\tilde w^2 - \tilde w_j^2)\,,\quad
\eta = C \prod_{i,j=1}^2(w^2- w_i^2)^2(\tilde w^2 - \tilde w_j^2)^2~.
\eeq
Note that on each of the $\IP^1$ there are just two points where
the elliptic fiber acquires a $D_4$ singularity: $w_i\sim -w_i,\,
i=1,2$ and $\tilde w_i\sim -\tilde w_j,\,j=1,2$. Each of them gives
rise to an $SO(8)$ enhanced gauge symmetry,
leading to a total $SO(8)^4$ gauge-group enhancement. There are now eight
points at which the $D_4$ singularities intersect:
\beq\la{inter}
(w_i,\tilde w_j)\,,\quad (w_i,-\tilde w_j)\,,\quad i,j=1,2~.
\eeq
This gives rise to eight tensor multiplets, and hence we have
$h_{1,1}=27$ and $h_{2,1}=3$. Note that the $Z_2$ action on the base has
restricted the $SL(2,\IC)\times SL(2,\IC)$ acting on the $\IP^1\times
\IP^1$ base to a $1+1$-parameter family rather than the full
$3+3$-parameter family. Thus, out of the six parameters
in~(\ref{etahztwo}), we are
still left with three parameters after rescaling~(\ref{weierfour}) and
using the above
restricted $SL(2,\IC)\times SL(2,\IC)$ action. From the discussion
above and using (\ref{nhyper}),
(\ref{tensorhype}) and (\ref{rankgg}), we find that in this
model $H_0=4$, $T=9$ and $r(V)=16$.
The gauge group $SO(8)^4$ gives rise to 112 vector multiplets. 
Inserting these values into (\ref{ganomaly}) we see
that the gravitational anomaly is apparently not satisfied.

This conflict raises the question whether the elliptic fibration
on the $Z_2\times Z_2$ orbifold $X_2$, with either the additional twist
in the Landau-Ginzburg representation or the additional shift
of (\ref{gammashift}), gives a consistent $F$-theory compactification.
However, it is observed~\footnote{We thank Andre Losev for
discussions
on this point.} that the action of the additional shift
(\ref{gammashift}) on the base coordinates commutes
with its action on the fiber. Thus, the additional shift
should still preserve the fibration. Hence, if the
fibration is consistent for the (51,3) Calabi--Yau
three--fold, it should still be consistent for the (27,3)
one, which is obtained from the (51,3) model by the
additional shift (\ref{gammashift}).
Note, however that the additional shift on the base
commutes with the shift on the fiber only for the 1/2
shift chosen. That is, any other shift $z_i\rightarrow z_i+a$
with $a\ne1/2$ will not preserve the fibration.
Furthermore, regarded as an orbifold of a flat torus, the $Z_2\times Z_2$
orbifold $X_2$ with $(h_{1,1},h_{2,1})=(27,3)$ is an orbifold
of a $T^6$ lattice, rather than $(T^2)^3$. However, as we
saw in the previous section, the $X_2$ model can be obtained
from the $X_1$ model by adding the freely-acting shift given in
(\ref{gammashift}), which preserves the cyclic permutation
of the $Z_2\times Z_2$ orbifold, and hence the factorization
into $({\tilde T}^2)^3$. This is consistent with the fact that
the (27,3) $Z_2\times Z_2$ orbifold model, for example in
its free-fermion realization, still possesses the cyclic
permutation symmetry between the three $T^2$ factors, which is
the characteristic property of the $Z_2\times Z_2$
orbifold. This factorization, and the existence of
the cyclic permutation symmetry, would naively suggest that
the (27,3) $Z_2\times Z_2$ orbifold model $X_2$ should still possess
a sensible elliptic fibration. 

But a general requirement of
a consistent $F$-theory compactification is that the elliptic fibration
produces
a global section. We will now show that this is not
fulfilled with the additional shift. 
Consider the quartic
representation (\ref{weierfour}) of the Weierstrass representation.
This quartic representation has two global sections, $\hat y=\pm\hat
x,\hat z=0$,
whereas that of (\ref{weier}) has only one, $y^2=x^3,z=0$. Under the
additional $Z_2$
action, the two sections in the quartic representation are simply
identified, except for at four fixed points in the base:
\beq\la{fixpts}
(w,\tilde w) = (0,0),\quad (0,\infty),\quad (\infty,0), \quad
(\infty,\infty)~.
\eeq
Note that because of the action on the fiber these are not fixed
points of the Calabi-Yau manifold, which remains smooth. 
Furthermore, at every point on
the base, other than the additional singular points of the fibration, the
transformation takes one point on the fiber to another point on the fiber.
However, at the singular points of the base
the fiber is shrunk to half 
its original size. The crucial observation~\footnote{We thank Paul Aspinwall
for pointing this out.} is that, whilst the intersection of a generic
fiber with the section is 1, it is 1/2 for the special fibers over the
fixed points. Thus, the section is not globally defined and $F$ theory
on $X_3$ is not well defined.

It is intriguing that this new puzzle
in the $F$-theory compactification on the (27,3) model
arises precisely because of the action of the
additional shift on the $T^2$ fiber.
To see how the above applies to the $X_2$ model, we consider
a related orbifold in which the $Z_2$ action is restricted to 
the base, leaving the elliptic fiber invariant.
In the examples of the previous sections, this corresponds
to the additional shift imposed on the $Z_2\times Z_2$ model
in the form (\ref{gammashift2}).
Taking the first two coordinates, $z_1$ and $z_2$,
to be the coordinates of the base and the third, $z_3$,
to be the coordinate of the fiber, we see that only the base
coordinates are identified under the additional shift, whereas the 
fiber is left invariant. In the case of this model,
as we saw in (\ref{alphagamma}) and (\ref{hodgez2z2shift4}),
there is an additional sector producing four additional $(1,1)$ and
$(2,1)$ forms. 
To see how these additional multiplets arise 
in the Weierstrass representation, note that
the additional $Z_2$ action is realized by
$(\hat y,\hat x, \hat z,w,\tilde w)\to (\hat y,\hat x, \hat
z,-w,-\tilde w)$. 
This gives the four fixed points in the base~(\ref{fixpts}),
and hence four fixed tori in the Calabi-Yau manifold,
as there is no action on the elliptic fiber. Each of these
tori contributes one K\"ahler form, from the two-cycle of the $\IP^1$
used to blow up the torus, and one complex structure deformation from
a three-cycle built of a family of $\IP^1$s over a one-cycle in the torus. 
Thus, $h_{1,1}$ and $h_{2,1}$ both increase by four. The rest of the 
analysis follows that of the previous model. 
After adding these contributions from the fixed tori, the spectrum is
$h_{1,1}=27+4=31,h_{2,1}=3+4=7$.

We see here the difference between the two models $X_2$ and $X_3$.
In $X_3$, the additional
shift (\ref{gammashift2}) is neither freely-acting on the base,
nor on the manifold regarded
as a Calabi--Yau three--fold, thus producing the additional
four $h_{1,1}$ and $h_{2,1}$ forms needed to resolve the singularity
in the conventional manner. However, in the $X_2$ model
the shift (\ref{gammashift}) does act freely on the
Calabi--Yau three--fold, although there are four fixed points in the base
$B$, and therefore does not produce any additional $h_{1,1}$ and $h_{2,1}$
forms associated with these singularities. 

We observe that
$F$ theory on $X_3$ has $T=13$ tensor multiplets, $H_0=8$ neutral
hyper--multiplets and $V=112$ vector multiplets, and
we see from (\ref{ganomaly}) that the
gravitational anomaly vanishes in this case.
Furthermore, from this example we see that the problem with the 
gravitational anomaly for the $X_3$ model arises strictly because
of the action of the additional shift (\ref{gammashift})
on the fiber. Thus, we would like to argue that a possible resolution of
the gravitational anomaly is the existence of one
hypermultiplet and one tensor multiplet at each of the fixed
points in the base.  
Still, how and whether the additional
singularities of the fibration may be resolved resulting in a
non-anomalous theory is 
an open question. However, it is tempting to
speculate that the resolution of these singularities
is intimately connected to the cancellation of the 
gravitational anomaly in this model. 

\setcounter{footnote}{0}
\section{Orientifold Representation}

In this section, we briefly discuss what would be the 
orientifold construction corresponding to
$F$-theory compactification on the $X_2$ orbifold.
Whilst it is not guaranteed that there exists an orientifold
model for every $F$-theory compactification, 
studying the orientifold construction may provide
a complementary way to understand physical issues.
In particular, in the case of the (27,3) model $X_2$,
the key question would be how to couple the additional
freely-acting shift to the orientifold projection.
In this connection, recall that, in this case, as the complex structure of
the fiber is identified with the dilaton of the
corresponding type-IIB string theory,
the shift in (\ref{gammashift}) acts non--trivially 
on the dilaton. 

We begin our brief discussion by studying
the orientifold corresponding to the
$F$-theory compactification of the $X_1$ orbifold~\cite{bz,dap,gm}.
We follow the analysis in~\cite{dap}, focusing in particular on the 
closed string sector as it seems to relevant for understanding the
missing tensor and hypermultiplets. Starting from $T^6$ given in terms
of complex coordinates $z_i,\,i=1,2,3$, and the identifications $z_i\sim
z_i+1\sim z_i + i,\, i=1,2,3$, the $Z_2\times Z_2$ action is given
by
\beqn\la{ztwoztwo}
&& Z_2^\alpha\,:\, (z_1,z_2,z_3) \to (-z_1,-z_2,z_3)\,,\\
&& Z_2^\beta\,:\, (z_1,z_2,z_3) \to (z_1,-z_2,-z_3)~.
\eeqn
We then let $z_3$ be the coordinate of the elliptic fiber, and define
the type-IIB orientifold on $T^4/Z_2^\alpha$ by the orientifold action
$\{1,\Omega(-1)^{F_L}R_2\}$. The $R_2$ acts on $T^4$ as $(z_1,z_2\to
(z_1,-z_2)$, and the remainder of the $Z_2^\beta$ action on the 
elliptic fiber is represented by $\Omega(-1)^{F_L}$~\cite{sen8d}. 

Clearly, in the absence of the $\Omega (-1)^{F_L}R_2$, we would just have
a type-IIB compactification on a $K3$ manifold, $T^4/Z_2^\alpha$, 
which gives an $N=2$ theory in six
dimensions with $T=21$ $N=2$ tensor multiplets, each of which
consists of one $N=1$ tensor multiplet and one $N=1$ hypermultiplet.
These arise from each of the sixteen fixed points of the
$T^4/\IZ^2$, as well as five 
from the untwisted sector. However, $\Omega(-1)^{F_L}R_2$ projects out
the hypermultiplets in the twisted sectors and leaves one tensor and
four hypermultiplets invariant in the untwisted sector~\cite{dap}.
Thus, we are left with $T=16+1$ tensor multiplets and $H_0=4$
hypermultiplets. This is the spectrum from the closed-string sector.
It can be shown~\cite{dap}\ by a more
careful analysis that there is an $SO(8)^8$ gauge enhancement
arising from the open-string sector, but no charged matter.

In order to understand how to implement the additional $Z_2$ action
required to get the 
$(h_{1,1}=27,h_{2,1}=3)$ $Z_2\times Z_2$ orbifold $X_2$
in the orientifold language, let us first consider a similar situation
in eight dimensions. 
It was shown by Witten~\cite{witten}
that there exists an orientifold in which there are three $O_+$
planes, one $O_-$ plane and eight
$D7$ branes. This is to be compared with the regular orientifold in
eight dimensions, in which there are four $O_+$ planes and 16
$D7$ branes. The latter corresponds to an $F$-theory compactification on
an elliptically-fibered $K3$ in which the monodromy of the elliptic
fiber is $SL(2,Z)$. By placing groups of four $D7$ branes
on each of the $O_+$ planes, cancelling the charge locally, one
obtains an $SO(8)^4$ gauge group~\cite{sen8d}. If
we do the same for the former model, we find a reduced gauge group
$SO(8)^2$, as there now only are eight $D7$ branes. In this case 
the $F$-theory compactification is an elliptically-fibered 
$K3$ with restricted monodromy $\Gamma_0(2)$~\cite{bps,bkmt}. Witten
argued that the existence of the two sets of orientifold planes is due
to a non-zero  flux through the NS-NS 2-form $B$, even if the 2-form
itself is projected out~\cite{witten}.

Let us now study the situation in six dimensions.  
 It was pointed out that just as in eight dimensions, by
turning on the NS-NS 2-form $B$, along one of the $T^2$ in the
underlying $T^4$, the rank is reduced by a factor of two~\cite{bpstwo}. 
In addition
the contribution from the closed string sector of the twisted sector
is changed. Rather than contributing 16 hypermultiplets, one instead
obtains 12 hypermultiplets and 4 tensor
multiplets~\cite{kst}. Since there is an
ambiguity in the action of the world-sheet parity
$\Omega$~\cite{polchinski}, we can consider a case with 4 $O_+$ planes
and 12 $O_-$ planes. Compared to the extreme case of 16 $O_-$ planes,
the gauge symmetry is reduced from $SO(8)^8$ to $SO(8)^4$, whilst there
are 12 tensor and 4 hypermultiplets. We recognize this as the spectrum
of the $F$-theory compactification of $X_3$. 
Thus, we find that it does not seem possible to construct an
orientifold with the properties corresponding to the $F$-theory based on
the $X_2$ orbifold.

We would like to remark that it is not at all clear that 
the orientifold construction will capture all of the non-perturbative
singularities  of the $F$-theory compactification.
In particular, in the case of the (27,3) $Z_2\times Z_2$
orbifold, the freely-acting shift (\ref{gammashift})
should act non-trivially on the dilaton multiplet.
Understanding the exact nature of this action
in the orientifold construction may provide
a complementary way to study the physical process involved.


\setcounter{footnote}{0}
\section{Discussion and Conclusions}

We have discussed in this paper $F$-theory compactification
on the Calabi--Yau three--fold which is associated 
with the realistic free-fermion models. 
The motivation to consider $F$-theory compactification on
this particular manifold is apparent: the realistic
free-fermion models have, after over a decade
of exploration, yielded the most realistic superstring models to date.
On the other hand,
over the last few years important progress has
been achieved in understanding non--perturbative aspects
of string theories. Although there is still a considerable way to go
before we have complete understanding what role non-perturbative string 
effects play in connection with  physics as observed in Nature,
some of the new tools have been applied to phenomenology. One such
example is the proposal~\cite{witten5d} to use
the eleventh dimension in $M$ theory to resolve the mismatch
between the unification scale calculated in the MSSM on the basis
of the values of the couplings observed at LEP and estimates
of the string unification scale in weak-coupling heterotic string
models~\cite{reviewsp}.

Among the gaps in our understanding of non-perturbative
aspects of string theory
is the ultimate mechanism that selects 
the string vacuum. One important aspect of this paper is that,
whilst the new puzzles that we have raised
are not well understood, {\it a priori} they may indicate
some non--trivial new physics associated with the dilaton multiplet
of the type-IIB string theory. We have exhibited and explored
elliptic fibrations corresponding to the orbifold models
$X_1$ and $X_3$. Although a six dimensional $N=1$ theory based on
the $F$-theory compactification of $X_2$ does not exist, we can 
speculate that the additional multiplets needed to
resolve the singularities, as well as the gravitational anomaly,
arise in some non--trivial non--perturbative way. The
most exciting may be the possibility of a still-unknown
physical phenomena that will provide a new view on
the dilaton fixing problem.  

\bigskip
\leftline{\large\bf Acknowledgments}
\medskip

We are pleased to thank Shyamoli Chaudhuri,
Lance Dixon, Tristan H\"ubsch,
Sheldon Katz, Peter Mayr, David Morrison, Ergin Sezgin
and especially Paul Aspinwall and Andre Losev for very helpful discussions.
This work was supported in part by the Department of Energy
under Grants No.\ DE-FG-05-86-ER-40272, DE-FG-03-95-ER-40917 and
DE-FG-02-94-ER-40823.
The work of P.B. was supported in part by the National
Science Foundation grant NSF  PHY94-07194. P.B. would also like to
thank the Aspen Center for Physics, LBL, Berkeley and TPI, Minneapolis
for hospitality during the course of this work.

\bigskip
\leftline{\large\bf Note Added}
\medskip

Subsequent to the submission of our paper, a Calabi-Yau
compactification with an elliptic fibration has been proposed
which has a very similar structure to the $X_2$ model discussed
here, namely a freely-acting shift with a bi--section and a
non--trivial $\pi_1$~\cite{ACK}.



\bibliographystyle{unsrt}

\end{document}